Impulsive Noise Immunity of Multidimensional Pulse Position Modulation

Michael D. Collins

Laboratory for Telecommunications Sciences, College Park, MD 20740

William B. Johnson

Laboratory for Physical Sciences, College Park, MD  20740

Abstract: We describe block oriented multidimensional pulse position modulation and its resilience against impulsive noise. The modulation implements the encoder and part of the decoder of the BBC algorithm.  We tested the modulation on circuits that send and detect a pulse based signal in the presence of impulsive noise.  We measured the packet error rate vs. signal to noise ratio and we compared it with published error rates for OFDM.  We found an error rate of $2 \times 10^{-5}$ at a signal to noise ratio of 16 dB without forward error correction and a data rate of 64 kbit/sec.

1. Introduction

   Modulation methods such as the orthogonal frequency division multiplex (OFDM) assume typically an additive white Gaussian noise (AWGN) model for the noise for calculating error rates. The probability of error for Gaussian noise is symmetric.  That is, after demodulation in the presence of Gaussian noise, it is equally likely that a message bit is flipped from a zero to one as from one to zero.  AWGN is convenient to use mathematically but it does not necessarily apply to the actual noise found in practice.  Impulsive noise, for example, does not follow the AWGN model.  Define a zero mark to be the absence of signal energy and a one mark to be the presence of signal energy.  Impulsive noise could typically add energy to the signal and therefore could change a zero mark into a one mark but it would not change a one mark to a zero mark. The impulsive noise would only make the amplitude of the one mark stronger.

   Multidimensional pulse position (MPP) modulation uses the encoder and part of the decoder from the Baird-Bahn-Collins (BBC ) algorithm (Baird, 2007) that specifically assumes the noise is totally asymmetric.  MPP is an energy pulse based system based on the descriptions in Baird, 2007.  In this paper, we define totally asymmetric noise to mean that the noise only flips a zero mark to a one mark but never flips one mark to a zero mark.  This is a strong assumption, but certain cases approximate this noise model, such as radar jamming and AC line noise.  In those cases, the BBC concurrent code allows for decoding of a signal in linear time as long as the received packet density is less than 50% of the total number of bits in a message. When the received packet density is greater than 50%, the time to decode the message goes up exponentially (Baird , 2007). We have not attempted to optimize the receiver in that regard.  In a fully implemented BBC pulse-based system, the marks created would likely be different and the sender message format much sparser. The receiver would be optimized to receive a packet density of 1/3.

To investigate the impulsive noise immunity of MPP, we tested MPP against the impulsive noise that is common on AC lines.  We chose to test the MPP on an AC line because impulsive noise usually dominates there over Gaussian noise.  The source of the impulsive noise is typically switched mode power supplies and electrical motors.  There is also cyclostationary noise that is frequency locked to the 60 Hz signal and consists mostly of 60 Hz and harmonics of 60 Hz caused by nonlinear response of rectifying diodes in power supplies and other electrical equipment.

We used custom circuits to implement and test the MPP.   We designed a simple pulse circuit that puts out a pulse with a width of a few microseconds and amplitude of 48 volts.  A simple high pass filter circuit couples the pulse onto the AC line and another high pass filter circuit couples the signal off of the AC line for measurement by an analog to digital converter (ADC).  The pulse circuit is controlled by a field programmable gate array (FPGA) which uses the BBC algorithm to encode a 64 bit ASCII message into 256 pulse positions in time.  The ADC circuit sends its output to another FPGA which decodes the received pulse positions back into an ASCII message which is compared with the original message to determine the packet error rate.

2. Multidimensional Pulse Position Modulation

Concurrent codes (Baird, 2007) were developed in 2007 at the USAF Academy and can be combined with a variety of modulation techniques such as MPP to provide high resilience to impulsive noise sources, e .g ., those encountered in radar jamming.

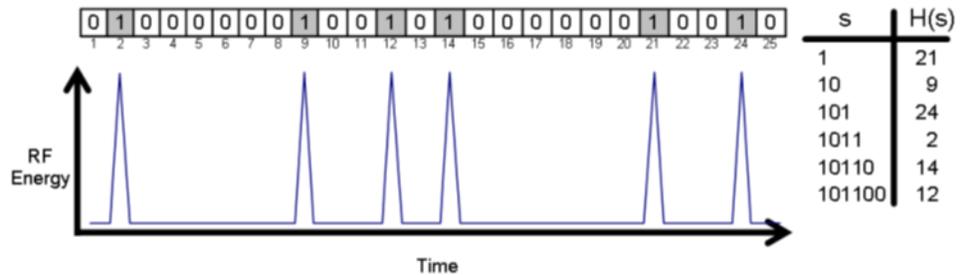

Figure 1:   Pulse positions for the BBC hash function H(s).  Taken from (Baird, 2007).

Figure 1 provides an example of multidimensional pulse position modulation.  In figure 1, the packet size is 24 bits and a single 6-bit message is encoded into the packet using BBC.  We want to emphasize we are encoding a single message or frame with a fixed bit size.  This results in a packet of 24 bits with 6 marks. This packet is sent in 24 time slots in which either a pulse is present or not present depending on the value of the bit in the reference packet. Furthermore, the position (index) of each of those marks within the packet is the hash value of a bit prefix of the 6-bit message (as shown in the table on the right of fig.1). This means that the positions are pseudo-random and interdependent. The overhead in this example is that six bits occupy 24 time slots so that, in this case, the data rate is 1/4 the rate of simple on/off keying.

We are making the fundamental assumption that none of the pulses are erased by noise spikes such as those that are common on an AC line. If a noise spike occurs in a position that already contains a pulse, then it is assumed to just increase the pulse amplitude. If a noise spike occurs in a position that does not contain a pulse, then with high probability our decoding algorithm still decodes the packet correctly. Extra pulses caused by noise spikes do increase the time to decode the messages.

If noise spikes increase the number of pulses, it is possible that a valid message, that was not actually sent, is decoded.  Such messages are called "hallucinations" (Baird, 2007). This issue can be controlled.   One method to do so is by including in the format of a message a trailing known-constant pattern.

We use the term "multidimensional" to distinguish MPP from the typical pulse position modulation, where each successive mark represents the next bits of the message and it goes into an additional time slot.   In MPP, on the other hand, where the marks occur depends on how the hash function acts on the bits in the message.   In fig. 1, the positions of all the marks could be completely different if the first bit of the message changes . The MPP in some sense has extra dimensions for the placement of the message bits.

a. MPP implementation

We implemented multidimensional pulse position modulation using the BBC encoding algorithm employing the GLOWWORM hash (Baird, 2012 and 2015). Our packet size is 256 bits and we encode a single message of 64 bits into a single packet using the GLOWWORM hash. Each time slot (i.e., pulse) is 3.9 µsec in duration. To encode a MPP packet, we start with a 64-bit message (8 characters of 8-bit ASCII), concatenate a trailing known-constant pattern of 13 zeros, and then map each of the 77 bits into a 256 bit encoded packet.  A packet is 1 msec (=256 x 3.9 µsec) long.   The data rate is then 64 (information) bits in 1 msec or 64 kbit/sec.  In the full BBC algorithm, the sender constructs a packet with much less than 1/3 of the bits set to 1, the receiver constructs a packet with exactly 1/3 of the bits set to 1, and the receiver decodes all messages in the packet.

To decode a MPP packet, an ADC captures a time series of 256-time slot positions of 3.9 µsec duration. This is done asynchronously. When a 3.9 µsec time slot is measured, a simple voltage threshold is applied to decide if there is a pulse in the time slot. When 256 time slot positions have been acquired, the sample is tested to determine if there is a valid message present. This means that 255 time slots after starting the detector, a full decode is performed on each and every time slot.

An important point is that if the central assumption of the BBC is violated and a pulse that was present in the original signal is erased by noise, the decoder will conclude incorrectly that no message is present and the decode will fail. That is, one such error means that the all data in that packet will be lost.

Due to design constraints, we do not implement the entire BBC decode algorithm.  We simply decode to the first valid code word received.  However, if we had decoded all valid code words, we would then have to include overhead in the packet to distinguish between messages we sent and hallucinations.  Since our packets are so small we chose to simply access the first code word returned by the algorithm.

If we had decoded all possible messages, then the measured packet error rate would have been smaller but the processing time would also have gone up.

A side effect of the MPP decoding is that all valid messages will be decoded and returned in lexicographical order.  Therefore, since the most significant bit of ASCII characters is 0, those messages should be decoded first until the number of hallucinations is too great. We control the received packet density by empirically adjusting power and the amplitude threshold. While not an optimal strategy for noise resilience, it performs remarkably well.

3. Circuit Implementation

   To test MPP we had to design a simple circuit that could send pulses on an AC line at specific time positions chosen according to the BBC hash function.

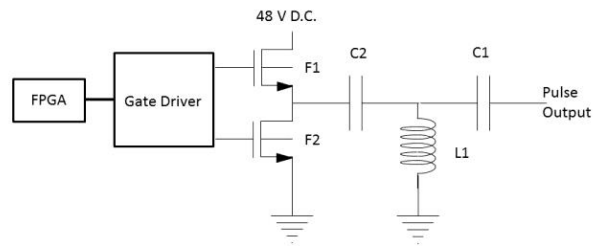

   Figure 2. A simple circuit used to output a 3.9 µsec pulse. L1 is 1 mH, and C1 and C2 are .47 µF high voltage capacitors. F1 and F2 are high voltage n-type MOSFET's. A gate driver is used to control F1 and F2 because F1 is floating off of ground.

   a.     Pulse Circuit

   Figure 2 shows the circuit that sends pulses. The AC line voltage drops across an LC high pass filter that consists of the capacitor C1 and inductor L1. The other side of the inductor L1 connects to the capacitor C2 which is charged by the FET F1 to a voltage of 48V and is discharged by FET F2. The FET's are both high voltage n-type MOSFET's (International Rectifier IRS4229PBF). Both FET's are controlled by an FPGA. A gate driver (Silicon Labs SiB2355BB-D-IM) is used to control the charging FET F1 since F1 is floating. In the operation of the circuit both F1 and F2 are used as on/off switches. When F1 is open, F2 is closed and when F2 is open, F2 is closed. F1 and F2 must never both close at the same time, since that will short the 48V D.C. power supply to ground. Note that the inductor L1 can be replaced by a transformer and the turns ratio of the transformer can be used to impedance match the pulse generator to the AC line. The self- inductance of one side of the transformer would form the L1 of a LC high pass filter with C1.

   The circuit generates a pulse in the following manner. F1 is closed and the capacitor C2 is charged up to the voltage of the 48 V D.C. power supply. F1 is then opened and F2 is closed.

The capacitor C2 then discharges a pulse thru the inductor L1. The pulse travels thru the capacitor C1 and out to the hot of the AC line. F2 is then opened and F1 is closed to recharge C2 and the cycle then repeats. Since an FPGA controls the charge and discharge of the MOSFET's, an arbitrary pulse sequence can be output.

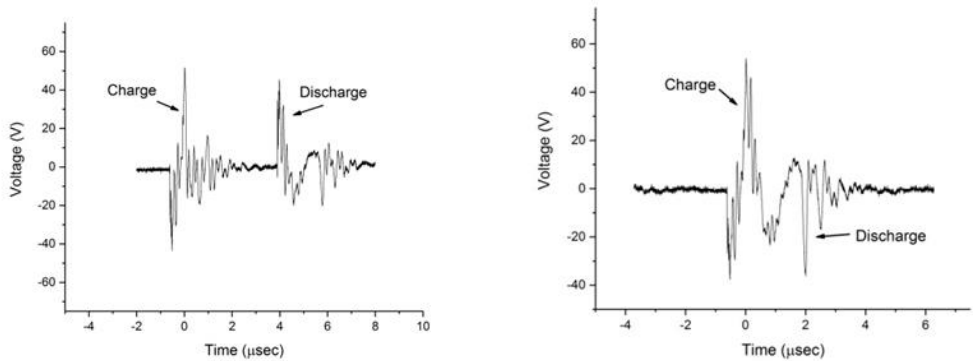

Figure 3. The pulses from the circuit of fig. 2. In the left graph, the discharge is 3.9 μsec after the charge and in the right graph, the discharge is .1 μsec later. The pulse on the right was used in the packet error measurements.

Fig. 3 shows the pulse waveform from the circuit of fig. 2. The amplitude of the pulse was 48 V. In fig. 3, the charging of capacitor C2 occurs first and then C2 discharges. The charging interval in fig. 3 is .5 μsec, and 3.9 μsec later, the discharge occurs for 1 μsec. There are two distinct pulses, first for the charge and then for the discharge. When the time between charge and the discharge decreases to .1 μsec, the charge and discharge pulses merge into one positive peak with ringing. The ringing in the pulse is caused by the inductor L1 and also by the inductance in the AC line. The overall duration of the pulse is controlled by the FPGA and varies from 3.9 to 10 microseconds. The amplitude of the pulse depends on the width of the charge cycle. For a charging time of .5 μsec, the positive peak is about 50 V and the negative peak is about 40 V. If the charging time is smaller than .2 μsec, then the pulse amplitude starts to decrease. The minimum width of the pulse peaks are set by the capacitance and switching speeds of the FET's F1 and F2. We used high voltage FET's for F1 and F2 and the pulse widths were around 1 microsecond. The ringing in the pulse limits the overall pulse width to 3.9 μsec or more. Hence, the fastest rate we can generate pulses is less than 250,000 pulses/sec. This is the ultimate limit on the data rate of this simple circuit as long as single amplitude pulses are used.

The pulse circuit fig. 2 generates a pulse between AC hot and ground, not between AC hot and neutral. We found that the pulse propagates farther if the pulse is between hot and ground rather than hot and neutral. We speculate that the reason is that many of the loads on a typical AC line are power supplies and motors that have a finite impedance between AC hot and neutral but not between AC hot and safety ground.  If a pulse sees a finite impedance load on a line, then some of the pulse current is lost into that load.  If the load is infinite (i.e., open) then the pulse does not lose current at that load.   By putting the pulse between AC hot and ground, less of the pulse current is lost as the pulse travels down the line and thus it travels farther.

    b. ADC circuit

To detect the pulses, we used a simple ADC circuit.  The circuit detected pulses on the AC line by sending them thru a high pass filter to strip off the AC line voltage and then a diode to clamp the voltage to +/- 10 V to protect the ADC.   A pulse then entered two op-amp's (Texas Instruments OPA335AIDBV and OPA301AIDBV ) which were used to both shift the signal to 0 - 5V and to provide either attenuation or gain to match voltage swing of the ADC.  The pulse was finally digitized by an ADC (Analog Devices ADG715 BRUZ).

We then used a voltage threshold to determine if a pulse was present in a given time slot so that it could be counted as a valid BBC mark.  Because of the presence of both positive and negative peaks in the pulse, we can use both an upper and a lower voltage threshold to determine whether to accept a pulse as a valid mark.  If a signal had either an amplitude above the upper voltage threshold level or below the lower voltage threshold, then the ADC accepted the signal as a valid mark. The use of the dual voltage threshold levels was crucial in mitigating the effect of 60 Hz noise (sec. 4)

4. Testing Procedure

    a. Noise

We tested the MPP against the AC noise in a laboratory.  Almost every plug in the laboratory fed either a desktop computer or laboratory equipment such as spectrum and network analyzers and voltage and laser sources and vacuum pumps.  This meant that almost every plug fed either a switching power supply or an electrical motor.

There were two dominant sources of noise in the AC line. One source was impulsive noise caused by switching power supplies and the other was noise caused by harmonics of the 60 Hz signal.   The AWGN noise was negligible.

To examine the noise in our laboratory, we used an Agilent DSO-X 6004 oscilloscope.  We could trigger the oscilloscope using the AC line voltage and we were thus able to resolve the noise into a part that was frequency and phase locked to the AC line and another part that was not locked.

The noise that triggered on the AC line had two parts. A low frequency part (180 Hz - 1 kHz) consisted of harmonics of the 60 Hz line signal. The amplitude was about 1.6 V. Another, higher frequency (7 kHz) impulsive part, also triggered on the AC line. The pulses were about 1 µsec in width with an amplitude around 2.2 V. A third source of noise was also impulsive, but it did not trigger on the AC line and was neither frequency nor phase locked to the 60 Hz signal. The third part consisted of pulses of about .5 µsec width with a frequency of 10 kHz and an amplitude of 3.2 V. These pulses came in bursts of 11 msec duration spaced 25 msec apart.

The impulsive noise satisfied the assumption of the BBC that no mark was erased: the impulsive noise did not erase a pulse in the original message. The effect of the impulsive noise was to add extra pulses in time slots that the original message did not have. As long as the received packet density was than 50% of the total time slots, our decoder algorithm decoded the original message in linear time. Otherwise, our decoder would output a hallucination, that is, the first valid ASCII message it found, which would appear as a string of random ASCII characters (i.e., gibberish) and not the original ASCII message.

The harmonics of 60 Hz could cause a pulse in the original message to fall below a simple voltage threshold. The 60 Hz harmonics are sine waves that have peaks and valleys and if a pulse fell in a valley of the sine wave it could then fall below a voltage threshold. This meant that a valid pulse would not be detected and this would violate our assumption that no valid mark/pulse is lost. In this case, our decoder fails completely to decode a valid message (sec.2). By using both an upper and lower voltage threshold, if the positive part of a pulse (fig. 3) falls below an upper voltage threshold and is not counted, then the negative part of the pulse might then fall below a lower threshold and be counted, so where the upper and lower thresholds are set is crucial to a successful decode.

We fit the probability density function (PDF) of our noise to a Middleton Class A model (Middleton, 1979). Before fitting, we normalized the noise by dividing it by the rms (root-mean-square) value of the voltage amplitude, so that the rms of the normalized noise was unity. The mean of the noise was zero. The fit had the parameters A= .305 and $\Gamma$= .046 (fig. 4). Middleton Class A model is not a good fit to narrowband PLC noise (Cortes, 2016), and that is the case here as well. We could either fit the peak at zero amplitude or fit the width of the peak, but not fit both. Our noise is certainly not Gaussian, since the plot of log(PDF) vs. amplitude is not parabolic, but has definite shoulders (fig. 4). An FFT of our noise showed definite peaks and not a continuous background, so our noise is not white either. The AWGN noise model thus does not apply. However, the Middleton model is used extensively (like the AWGN model) in simulations for OFDM with which we compare our results (section 5b), so we include it here.

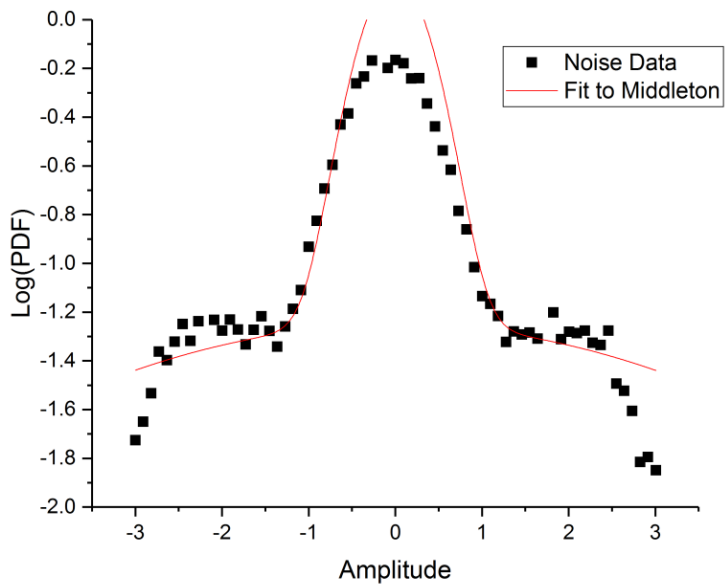

Figure 4. Fit of the voltage noise in our experiment to the Middleton Class A model. The x-axis is the normalized amplitude of the noise and the y-axis is the common logarithm of the probability density function of the noise. We fit to three terms in the Middleton Class A model.

b. Threshold Level

Upper and lower thresholds were set by the following procedure. Packets are sent on an AC line and detected and counted by an ADC and sent to an FPGA and then to a computer monitor. The ADC signal varied numerically between 0 and 4000. The 4000 level corresponded to 4 V, the zero level corresponded to -4V and a level of 2000 corresponded to zero volts (i.e., ground voltage).

To set the thresholds, first the upper threshold was set to 4000 and the lower threshold was set to zero, so that no pulses were accepted as valid and there were no decodes. The upper threshold was then lowered from the value of 4000 by increments of 100 until valid decodes of the ASCII message began. We then continued to lower the upper threshold until so many extra noise pulses were detected that our decoder on the FPGA put out random ASCII strings (i.e., gibberish). The upper threshold was then increased by 100 and fixed. The lower threshold was then increased from zero by steps of 100 until again ASCII gibberish was output. The lower threshold was then decreased by 100 and fixed. We then recorded 10 minutes of data and counted how many valid messages were received. The percentage of dropped ASCII messages out of the total number of messages sent was recorded as the packet error rate.

c. Signal/Noise Ratio

We measured the signal to noise ratio by sampling the pulse waveform out of the detection circuit with an Agilent DSO-X 6004 oscilloscope. We used a Tektronix 6104 10x probe to observe the high voltages without damaging the oscilloscope. We observed both the signal and the noise after they passed through a high pass filter that removed the 60 Hz line signal. The high pass filter we used was the same as the one used in the pulse circuit; i.e., the filter formed by C1, C2 and L1 (fig. 2).

To compare our results with simulations of OFDM we plot our packet error rate vs. the signal to noise ratio $Eb/Nb$. $Eb$ is the energy per bit. It is proportional to the rms value of the signal voltage. $Nb$ is the noise energy per bit with units of e.g., joules. It is proportional to the rms value of the noise voltage. A typical error rate plot uses $N0$ (power spectral density in units of e.g., joules/sec Hz) and plots error rate vs. $Eb/N0$. $N0$ makes sense for white noise where $N0$ is constant vs. frequency, but it does not make sense for impulsive noise where the $N0$ varies drastically with frequency. The noise energy per bit $Nb = N0*W*B$ where W is the bandwidth of the detector and B is the bit period. We get Nb by taking a scope trace (200 ms) that is long with respect to a bit period (16 µsec) and averaging the rms value of the noise over 5000 scope traces. The 16 µsec bit period occurs because one MPP bit has four pulses of 4 µsec per pulse. To find the ratio $Eb/Nb$, we took the ratio of their respective rms voltages since all the other factors in Eb and Nb were the same and canceled out.

To find Eb, we used the oscilloscope to measure both the maximum positive voltage and the rms voltage of a pulse much larger than the noise and found the ratio of the two voltages. We used the oscilloscope to find the maximum value of the pulse for pulses of various sizes as long as the maximum of the pulse was larger than the maximum of the noise. We then used the ratio of rms to maximum measured for the large pulse to find the rms for the smaller pulse. This avoided the problem that recording the rms for the smaller pulse would include both the rms of the pulse and the rms of the noise. We recorded the mean value of the maximum over multiple traces. We also recorded the rms noise when no pulse was present as described in the previous paragraph. In practice, we accumulated statistics for about 5000 oscilloscope traces.

When the pulse was smaller than the noise, we could not use this procedure because the maximum of the pulse was less than the maximum of the noise. Instead, we recorded a trace and picked out the pulse and measured the maximum voltage by using the cursors on the oscilloscope. This was less precise than gathering statistical data, but it was unavoidable.

We can relate published simulated OFDM error rate curves (Shongwe, 2015) to our curves since for an optimal OFDM detector B*W is about 1. Optimal means that the detector has just enough bandwidth to detect the signal but not so much that it lets in extra noise. Thus, if the bandwidth is much less than 1/B, the detector will distort the bits. If W is much greater than 1/B, the detector is letting in too much noise. So an optimal OFDM detector will have B*W=1. That means that Eb/Nb = Eb/(N0*W*B) = Eb/N0 and we can compare the published OFDM simulation curves with our curve.

   d. Packet Error Rate

We measured the error rate in the MPP decodes by the following procedure. We sent messages that consisted of 8 packets of 256 bits per packet. Each packet was 1 ms in duration and contained the MPP coding of an 64 byte ASCII word such as "Hello1!\" where the ASCII coding was 8 bits per ASCII character and the "\" was a line feed. Each of the eight packets encoded a different ASCII word. The sequence of words was the same in each message.

The messages were 8 ms long, which we chose because it is 1/2 the 16 ms period of 60 Hz. We wanted to measure a worst case error rate and an 8 msec message length increased the chances that 60 Hz harmonic interference would push a packet below the threshold level and cause it not to be detected and thus cause an error. Between messages, we inserted a 50 ms pause so that we could print the received messages on a computer monitor as well as store the messages in a file for later processing.

To measure the packet error rate at a given signal/noise ratio, we repeated the same message about 12,000 times over a period of about 10 minutes. We wrote a computer program to count the number of successfully detected packets and also the number of completely corrupted packets. The corrupted packets contain none of the original ASCII words, but only random ASCII characters. There were also messages that did not contain all the sent packets, which meant that they were missing some of the ASCII words. We counted the number of correctly received packets for each of the 8 packets in each message and calculated the average number of successfully received packets for each of the 8 packets. Each packet encoded one of our ASCII words. The total number of packets sent was 8 x number of messages, which was about 8 x 12,000, or 96,000. The number of dropped packets was this total minus the number of successfully received packets. The packet error rate was the number of dropped packets/total and is shown in fig. 5.

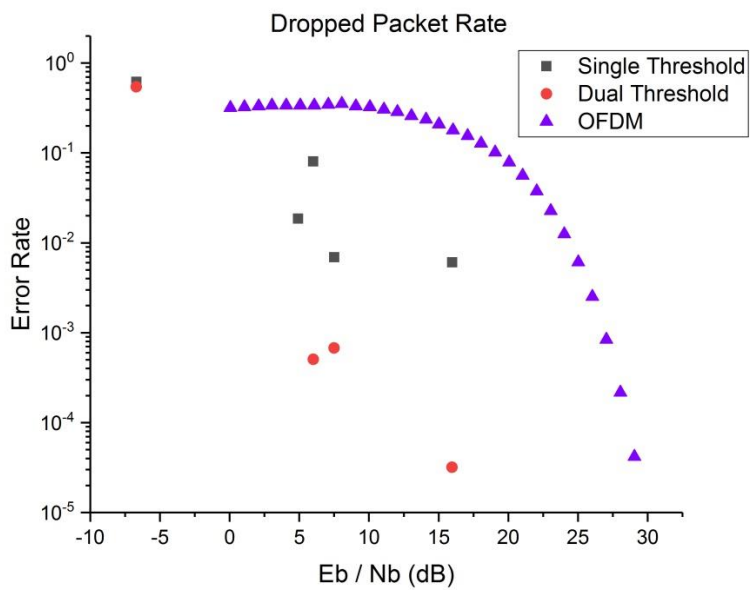

Figure 5: The measured MPP packet error rate vs. signal/noise ratio measured for the pulse circuit of fig. 2. Most of the noise was impulsive noise on an AC line but some was harmonics of 60 Hz. The Middleton Class A parameters for our noise are A = .305 and Γ = .0455. The OFDM data is for comparison and is from fig. 2 of (Shongwe, 2015). The QPSK-OFDM data is a simulated bit error rate for random impulsive noise with Middleton parameters A = .1 and Γ = .01 with no preprocessing or error correction. The MPP information data rate is 64 kbit/sec.

5. Discussion

    a. Comparison of Single and Dual Thresholds

    There are data points in fig. 5 for both the single (upper voltage) threshold curves and for dual (both upper and lower voltage) threshold curves (sec. 4b). The pulses that were used in the experiment to represent the bits have both a positive voltage peak and an adjacent negative voltage peak (fig. 3). The two peaks were artifacts of the circuit used to generate the pulses. When a single threshold was used, an ADC accepted packets as valid when the positive peaks were above a preset voltage. When a dual threshold was used, the ADC accepted packets as valid when either a positive pulse peak was above a preset upper voltage threshold or a negative pulse was below a lower preset voltage threshold. The dual threshold curve (fig. 5) is one to two orders of magnitude lower in packet error rate than for the single threshold curve. We attribute this to the fact that the lower threshold counts pulses that would otherwise be lost when the 60 Hz noise pushes a pulse below the upper threshold (sec. 4) but below the lower threshold.

b.  Comparison with OFDM

OFDM is widely used, for example, for modulation on AC lines (Rindchen, 2015) and for 802.11 wireless (Banerji, 2013 and Gast, 2005).  OFDM features a high data rate because it uses parallel orthogonal frequency channels and has immunity to both AWGN and other kinds of noise when combined with forward error correction (FEC).  However, it is not very good against impulsive noise (Shongwe, 2015).  The MPP modulation that we tested in this paper was specifically designed for immunity to impulsive noise and seems to perform better than OFDM for impulsive noise.  To demonstrate this, we compare our data with published error rate curves for an OFDM simulation with impulsive noise (figures 2, 8, and 9 of Shongwe).  We compare with simulated QPSK-OFDM bit error rate curves with FFT size N = 2048 that used a Middleton Class A noise model with parameters A = .1 and $\Gamma$ = .01.  This OFDM simulation was as close as we could find to the Middleton Class A fit of our data with parameters A = .305 and $\Gamma$= .0455.  We compare our error rate against a simulated OFDM curve that had no error correction or preprocessing (fig. 2 of Shongwe, graph labelled as random impulsive noise: no preprocessing) and curves that had error correction and preprocessing (Shongwe, fig. 8 and 9 ) .  We find that the MPP is at least two orders of magnitude lower error rate than OFDM against random impulsive noise for Eb/Nb (=Eb/N0)  ratios between 0 dB and 16 dB when we used the dual threshold method (fig. 5).  For Eb/Nb ratios less than unity, the MPP and the OFDM error rate were the same within an order of magnitude.  Note that preprocessing and error correction in the OFDM simulation (Shongwe, 2015) could bring the error rate down to close to the AWGN limit for other parameters of the Middleton noise model and this performance is better than our MPP result which has no error correction.

6. Conclusions

We measured the packet error rate vs. Eb/Nb for physical circuits that used the multidimensional pulse position (MPP) modulation.   We described both the MPP (sec. 2) and the physical circuits (sec. 3) used to send and detect MPP pulses.   We tested the MPP using an AC line that contained impulsive noise and 60 Hz harmonic noise and negligible AWGN.   We discussed two methods for setting a voltage threshold on an ADC that detects the pulses (sec. 4).  Using dual thresholds, an upper and a lower threshold, rather than a single threshold, lowers the packet error rate vs. Eb/Nb by two orders of magnitude (sec. 5).  The multidimensional pulse position modulation method with dual threshold achieves a packet error rate of around $2 \, x \, 10^{-5}$ at an Eb/Nb  of 16 dB with no FEC (fig. 5).  This is better than published BER curves for OFDM with random impulsive noise with no error correction by two orders of magnitude although the published OFDM curves were for somewhat different noise parameters.

Acknowledgements:

	We wish to thank the following individuals who worked on this project:  Laboratory for Telecommunication Sciences: Dr. Jon Kosloski, Dr. Andrew Culhane, Mr. Nathan Thai, and Dr. Manishika Agaskar for software development.   Signalscape, Inc. : Mr. Mark Kalendra, Mr. Matthew Purvis, and Mr. Mike Lenzo  for hardware and software development.  Laboratory for Physical Sciences: Dr. Thomas Mielke for help with testing.  ICASA:  Mr. Michael Smith, Mr. Alexander  George, and Mr. Daniel Ericson for hardware development and propagation studies.